\documentclass[epjCONF]{svjour}
\usepackage{graphics}
\usepackage[varg]{txfonts} 
\usepackage[latin1]{inputenc}
\session-title{CNR*09}
\begin{document}
\title{Extending MCAS to hypernuclei and radiative-capture reactions}
\author{L. Canton\inst{1}\fnmsep\thanks{\email{luciano.canton@pd.infn.it}} 
\and K. Amos\inst{2} 
\and S. Karataglidis\inst{3}
\and J. P. Svenne\inst{4}}

\institute{Istituto Nazionale di Fisica Nucleare, sezione di Padova,
via Marzolo 8, Padova I-35131, Italia \and 
School of Physics, University of Melbourne, Victoria 3010,
Australia \and 
Department of Physics and Electronics, Rhodes University,
Grahamstown 6140, South Africa \and
Department of Physics and Astronomy, University of Manitoba,
Winnipeg, MB, Canada R3T 2N2}
\abstract{ Using a Multi-Channel Algebraic Scattering (MCAS) approach
we have analyzed the spectra of two hypernuclear
systems,  $^9_\Lambda$Be  and $^{13}_\Lambda$C. We have studied the
splitting of the two odd-parity   excited levels (${1}/{2}^-$  and 
${3}/{2}^-$)  at  11  MeV   excitation    in $^{13}_\Lambda$C,
originated by  the weak $\Lambda$-nucleus
spin-orbit force. We have also considered the  splittings  of  the 
${3}/{2}^+$  and ${5}/{2}^+$ levels in  both $^{9}_\Lambda$Be
and $^{13}_\Lambda$C, finding how they originate from 
couplings to the collective  $2^+$
states of the core nuclei. In both hypernuclei, we suggest that there 
could be additional  low-lying  resonant states  
in  the $\Lambda$-nucleus continua. From the MCAS approach one
can extract also the full coupled-channel  scattering wavefunction to be
used in the calculation of various  transition matrix elements. As a
first application, we have considered the  EM-transition matrix elements
for the capture reaction $\alpha$ + $^3$He $\rightarrow$ $^7$Be + $\gamma$.
} 
\maketitle

\section{Introduction}
\label{intro}
In recent years, a Multi-Channel Algebraic Scattering (MCAS) approach has
been developed to describe  low-energy nucleon-nucleus scattering, 
resonance phenomena, and sub-threshold spectra for medium-light nuclei.
We have considered compound nuclei that are stable, or weakly bound,  
and have also extended the analysis to very unstable systems that are 
unbound with respect to proton emission, namely, that are  beyond the
proton drip line.

In this paper we illustrate first  the possible MCAS description  of the
levels of two hypernuclei $^9_\Lambda $Be and $^{13}_\Lambda $C. These
systems are described in terms of a phenomenological  $\Lambda$-light
mass nucleus interaction which explicitly couples the hyperon to the
collective low-lying states of the ordinary nuclear core.  The
phenomenological  character of an appropriate $\Lambda$-light mass
nucleus interaction was   established in recent reviews of hypernuclear
theory and experiments~\cite{Mi07,Ha06,Ta08}. Essentially it has a
central depth of $\sim 30$ MeV (with a  Woods-Saxon form) and a
spin-orbit attribute considerably weaker than that of a nucleon-nucleus
interaction. 

The MCAS approach to particle-nucleus structure and low-energy
scattering~\cite{Am03} emphasises the couplings of
single-particle dynamics  with  low-lying collective excitations  of the
ordinary nuclear core. The current role of these studies is to analyze
bound and resonant spectra to support and interpret experimental
investigations. When the Pauli principle is
incorporated in nucleon-nucleus dynamics, the method has been  shown to
describe, consistently, the bound and resonant spectra of normal
(zero-strangeness) light-mass nuclei~\cite{Am03,Ca06,Ca06a}. 
Starting with the properties of spectra of non-strange nuclei, we 
consider the modifications to the Hamiltonians in MCAS that are  
required  to describe hyperon-nucleus dynamics.  In particular, we
analyze low-lying level structures of two  $p$-shell $\Lambda$
hypernuclei with regards to the structure of the hypernuclear doublet
levels.  We consider splittings that have been measured recently and, as
well,  find level structures that  are just above the $\Lambda$-nucleus
scattering threshold. Perhaps they may be observed in future
experiments.

As a second application of the MCAS method, we consider 
the ${^{3}{\rm He}}({\alpha},\gamma ){^{7}{\rm Be}}$ reaction cross section
at astrophysical energies. The model Hamiltonian is based on a 
simplified 2-cluster model. The energy-dependence of the calculated 
astrophysical factor reproduces the experimental situation fairly well, 
though cross-section normalization turns out to be slightly overestimated.

\section{
The two hypernuclei $^{9}_\Lambda $Be and $^{13}_\Lambda $C}
\label{sec:1}

\subsection{
The $\Lambda$-$^8$Be interacting system}
\label{Be results}
Recent works on $^9_\Lambda$Be concentrate on the 
ground level and on the splitting of the two first excited levels
$\frac{5}{2}^+$ and $\frac{3}{2}^+$ \cite{Mi07,Ta08,Ta05a,Ak02}.
These states are all of positive parity and 
herein we also consider those  positive-parity states.
Extensive theoretical work has been done
in the past on low-energy negative-parity states for this system,
but the relevant details are omitted herein for brevity. 

To describe the $\Lambda$-$^8$Be  system we applied
scalings to a previously determined $n$-$^8$Be Hamiltonian, 
adjusting the  parameter values by
reducing the central potential by 30\%,   reducing the interaction radius
by 15\%, and reducing the spin-orbit strength by an order of magnitude.
In dealing with $n$-$^8$Be, an appropriate OPP term had been
included in the nucleon-core Hamiltonian~\cite{Ca05},
since the deeply bound $s$ states are
already occupied by the core nucleons.
However, when considering the hyperon-nucleus system,
these OPP terms are removed since the hyperon is
distinguishable from the nucleons and access to the
deeply-bound orbits is not suppressed by Pauli exclusion.

Small adjustments were then made to give
best results. The final values of the  coupled-channel potential
parameters are given in Table~\ref{table3Lambda}. Therein, 
we consider two sets of parameters, one with the onset of a small
spin-spin component of the interaction (`Case 2').
\begin{table}[ht]
\begin{center}

\caption
{\label{table3Lambda} Strengths (in MeV) of the $\Lambda$-${}^{8}$Be interaction
with $R = 2.3$ fm., $a = 0.65$ fm., and $\beta_2 = 0.7$}
\begin{tabular}{|ccc|}
\hline\hline
  & {Case 1}  & {Case 2} \\
\hline
$V_0$ & $-$26.4 & $-$26.4\\
$V_{\ell s}$ & 0.35 & 0.35 \\
$V_{sI}$ & 0.0 & $-$0.1 \\
\hline\hline
\end{tabular}
\end{center}
\end{table}
In Table~\ref{Table1-BeLam}, we give the spectrum calculated with the
MCAS approach and the two sets of parameters given in  
Table~\ref{table3Lambda}. We report the experimental  binding energy of
$^9_\Lambda$Be as determined from emulsion data~\cite{Ju73}, but have
also indicated in brackets the additional, newer value obtained by the
E336 experiment at KEK (see Ref.~\cite{Ha06}). While the emulsion data
for $^7_\Lambda$Li and $^{13}_\Lambda$C agreed well with the
KEK results, there was significant disagreement between those 
regarding the $^9_\Lambda $Be ground-state binding energy, with   the
reason  for this disagreement not known.    Also, we consider for
comparison shell-model results. These results \cite{mi05} were obtained
by considering not only the $\Lambda N$ and $\Sigma N$ interactions but
also the $\Lambda \Sigma$ coupling. The $\Lambda$ was assumed to be in
the $0s$ orbit while the nucleons were assumed to be in the $0p$-shell.
Comparison with the bound-state spectrum obtained from MCAS is quite
good.

\begin{table}[ht]
\begin{center}
\caption{
The positive-parity spectrum, in MeV, of ${}^9_\Lambda $Be
(in round brackets we give the resonance widths in keV).  
The columns labelled `Case 1' and `Case 2' 
refer, respectively,  to calculations made without and with the 
spin-spin term in the potential.  The spin-spin strength, $V_{s\cdot I}$,
was $-$0.1 MeV.  Comparison is
also made with the results of a shell model (SM) calculation \cite{mi05}, where the ground state binding
energy has been set to the measured value for comparison.
\label{Table1-BeLam}}
\begin{tabular}{|ccccc|}
\hline\hline
$J^\pi$  & Exp.~\cite{Ak02} & Case 1 & Case 2 & SM~\cite{mi05} \\
\hline
$\frac{7}{2}^+$   &  $--$   &     4.791 (4.1) &  4.92 (4.9)  &    \\
$\frac{9}{2}^+$   &  $--$   &     4.788  (4.4) & 4.68 (3.8)   &  \\
$\frac{3}{2}^+$   &  $-$3.64  &   $-$3.70  &   $-$3.63  &  $-$3.66        \\
$\frac{5}{2}^+$   &  $-$3.69  &   $-$3.65  &   $-$3.70   & $-$3.71        \\
$\frac{1}{2}^+$   &  $-$6.71 &   $-$6.73 &    $-$6.73   & $-$6.71  \\
\hline\hline
\end{tabular}
\end{center}
\end{table}

It is interesting to observe that we get the correct size of
this fine splitting between these two states with a simple phenomenological
model consisting only of a central and a spin-orbit potential
(`Case 1' in Table~\ref{table3Lambda}). Indeed, 
assuming no $sI$ coupling,
the  magnitude of the  splitting between the $\frac{3}{2}^+$ and
$\frac{5}{2}^+$ states is very small, but 
consistent with the separation value recently measured~\cite{Ta05}. 
But the $\frac{3}{2}^+$ state is predicted to be the lower, at
variance with the recent analysis~\cite{Ta05a}, where 
the $\frac{3}{2}^+$ state was assessed to be the less bound of the pair.

To achieve the correct level ordering requires the introduction of a small
spin-spin contribution in the $\Lambda$-$^8$Be phenomenological
interaction (`Case 2'). We could not obtain the observed level ordering
without that component in the interaction.

At 4.7 MeV above the scattering threshold, we predict two additional
positive-parity states (resonances).
These are formed by  coupling the
4$^+$ state in ${}^8$Be with a $\Lambda $ in the $s_\frac{1}{2}$ state.
The widths of these resonances shown in brackets in
Table~\ref{Table1-BeLam}
were calculated assuming that the
${}^8$Be-core 4$^+$ state has zero width. If we consider
the $\alpha$-decay probability of this $4^+$ level,
then the widths reported in the table can be expected to increase quite
significantly~\cite{Fr08}.

\subsection{
The $\Lambda$-$^{12}$C interacting system}
\label{C results}

The $^{13}$C (and ${}^{13}$N) systems have been studied extensively
using the MCAS approach~\cite{Am03,Ca05,Ca07,Sv06}. 
Good reproduction of the
low-lying spectra, sub-threshold and resonant states,   and of the
elastic scattering cross-section and polarization data have been
obtained with a relatively simple model Hamiltonian.  

                Here, we consider application of MCAS for the
$\Lambda$-$^{12}$C system.   Similarly to the previous ${}^9_\Lambda$Be
case,   we start with the  depth of the  $\Lambda$-nucleus optical
potential  which is is about $\frac{2}{3}$ that of the standard
nucleon-nucleus one.     The spin-orbit strength also is much smaller,
an order of magnitude smaller than that of the corresponding
nucleon-nucleus system. Additionally, the potential radius is 
$\sim$15\% smaller than used for the $n$+${}^{12}$C system. Starting from
that set of parameter values, we have tuned the coupled-channel
potential interaction to the description of the known spectrum.  The
resultant final potential parameters we have used are listed in
Table~\ref{table1}.
\begin{table}[ht]
\begin{center}

\caption
{\label{table1} Strengths of the $\Lambda$-${}^{12}$C interaction
with $R = 2.6$ fm., $a = 0.6$ fm., and $\beta_2 = $-$0.52$ }
\begin{tabular}{|ccccc|}
\hline\hline
  &\multicolumn{2}{c}{Case 1}  &\multicolumn{2}{c}{Case 2} \\
  & $\pi = -1$ & $\pi = +1$ & $\pi = -1$ & $\pi = +1$\\
\hline
$V_0$ (MeV) & $-$28.9 & $-$30.4 & $-$28.9 & $-$30.4\\
$V_{\ell s}$ (MeV) & 0.35 & 0.35 & 0.35 & 0.35 \\
$V_{sI}$ (MeV) & 0.0 & 0.0 & $-$0.1 & $-$0.1 \\
\hline\hline
\end{tabular}

\end{center}
\end{table}
Again, the parameters identified as `Case 1' were those we found with
the spin-spin interaction strength set to zero,
while those defined as `Case 2' had
the small spin-spin interaction strength listed.
The parities of the channel interactions are designated by $\pi = \pm 1$.

\begin{table}[ht]
\begin{center}
\caption{Spectra of
$^{13}_\Lambda$C with energies in MeV. In round brackets we denote
the corresponding widths (expressed in  keV when using ``k", 
otherwise in MeV).  
The shell model results are those from Ref.~\cite{Mi09}.
\label{table2}}
\begin{tabular}{|ccllc|}
\hline\hline
$J^\pi$  & Exp.~\cite{Ko02} & Case 1 & Case 2 & SM~\cite{Mi09} \\
\hline
$\frac{1}{2}^-$   &  $---$    &   +4.65  (0.21)  &   +4.66 (0.23) & \\
$\frac{3}{2}^-$   &  $---$    &   +4.64  (0.22)  &   +4.63 (0.21) & \\
$\frac{5}{2}^-$   &  $---$    &   +4.28  (1.0k)   &   +4.31  (1.0k) & \\
$\frac{7}{2}^-$   &  $---$    &   +4.17  (1.0k)   &   +4.14  (1.0k) & \\
$\frac{3}{2}^-$   &  $---$    &   +3.10  (0.1k)   &   +3.15  (0.1k) & \\
$\frac{5}{2}^-$   &  $---$    & +3.05 (0.1k)   & +3.02 (0.1k) & \\
$\frac{1}{2}^-$   &  $-$0.708 &    $-$0.74\          &   $-$0.74\      &   \\
$\frac{3}{2}^-$   &  $-$0.86\ &    $-$0.89\          &   $-$0.89\      &   \\
$\frac{1}{2}^+$   &  $---$    &    $-$4.12\          &   $-$4.12\       &  \\
$\frac{3}{2}^+$   &  $-$6.81\ &    $-$7.177          &   $-$7.08\   &   $-$6.22   \\
$\frac{5}{2}^+$   &  $-$6.81\ &    $-$7.178          &   $-$7.24\   &  $-$6.19    \\
$\frac{1}{2}^+$   &  $-$11.69 &   $-$11.68           &   $-$11.68  & $-$10.95        \\
\hline
\end{tabular}
\end{center}
\end{table}

Four state energies  have been observed and they are
listed in Table~\ref{table2}
in  the  column  labelled  `Exp'.  A splitting  between  the
$\frac{3}{2}^+$ and $\frac{5}{2}^+$ states is not resolved as yet.   The
theoretical spectra, however, contain a richer  structure as shown by
the results listed under the `Case 1' and `Case 2' columns in the table.
Both model calculations predict a $\frac{1}{2}^+$ bound state  at 4.12
MeV below threshold. As the spin-spin interaction has no effect upon its
excitation energy, this state corresponds to an
$s_{\frac{1}{2}}$-$\Lambda$ coupled to the $0_2^+$ state at 7.65 MeV
in $^{12}$C, which is an highly exotic state as it corresponds to the 
coupling of the hyperon to the superdeformed Hoyle state. This state is not
predicted by the shell model \cite{Mi09}, as the $0p$-shell model of the underlying
structure of $^{12}$C cannot predict the Hoyle state. The ground state and low-lying spectrum from
the shell model for $^{13}_\Lambda$C, however, agree generally well with the predictions from MCAS.
The $\frac{1}{2}^+$ state we expect at 7.56 MeV excitation in
$^{13}_\Lambda $C has been placed much higher in excitation (12.2 MeV)
in the cluster model evaluations~\cite{Hi00}. That is due to the strong
state dependence of the  $\Lambda$-nucleus interaction, found by
$s$-wave folding in that model~\cite{Ya85}.
We have still to wait for more detailed experimental 
investigations of that spectrum
to decide whether it is a strong state dependence effect or simply
the extremely weak natural excitation of the second $0^+$ from the
ground state of $^{12}$C that explains the position of a second
$\frac{1}{2}^+$ state in the spectrum of this hypernucleus.
Note that a similar
state has been predicted by MCAS calculations of $n$+${}^{12}$C to be
in the spectrum of the non-strange $^{13}$C.
That state has spin-parity $J^\pi=\frac{1}{2}^-$, lies 2.68 MeV
above the scattering threshold,  and though unobserved, is partner to
a mirror state  of such structure  in $^{13}$N at
6.97 MeV above the $p$-${}^{12}$C threshold. That is
1.3 MeV higher than where a $J^\pi=\frac{1}{2}^-$ state has been
observed in $^{13}$N.

Additionally,
a set of six odd-parity states of $^{13}_\Lambda $C are predicted to be
just a  few MeV above the scattering threshold.
They are states formed by coupling of a $p_{\frac{3}{2}}$- and of a
$p_{\frac{1}{2}}$-$\Lambda$ with the $2^+_1$ state in ${}^{12}$C.
Without deformation (and no spin-spin interaction) they would form
a degenerate
quartet ($\frac{1}{2}^-, \frac{3}{2}^-, \frac{5}{2}^-$, and  $\frac{7}{2}^-$)
and a degenerate doublet ($\frac{3}{2}^-$ and  $\frac{5}{2}^-$) of states.
Deformation and spin-spin effects then break those degeneracies and
mix states of like spin-parity.  Since these six states are
embedded in the $\Lambda$-$^{12}$C continuum, they are resonances
and their widths are listed (in brackets) in Table~\ref{table2}.
These resonance states are very narrow save
for the doublet of $\frac{3}{2}^-\vert_3$ and $\frac{1}{2}^-\vert_2$
resonances.  We have also calculated  the
corresponding excitation function, assuming a low-energy $\Lambda$-$^{12}$C
elastic scattering process.

\begin{figure}[h]
\scalebox{0.45}{\includegraphics*{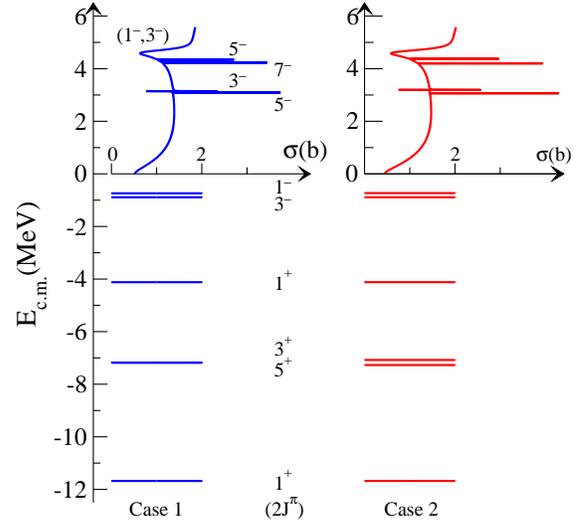}}
\caption{\label{Canton-fig2}
The spectra and total elastic cross sections for
$\Lambda$-$^{12}$C scattering resulting from MCAS calculations
made using two sets of parameter values.}
\end{figure}
The spectra for $^{13}_\Lambda $C is given in
Fig.~\ref{Canton-fig2}. Below the $\Lambda-^{12}$C threshold
the discrete bound states are shown while above that threshold the
theoretical cross sections are given for both the case of $V_{sI}$ =
0.0 and $-$0.1 MeV.
Of prime interest is the fine splitting of the bound  $\frac{1}{2}^-$  and
$\frac{3}{2}^-$ levels,   observed respectively at 10.98 MeV and 10.83 MeV
above the ground state~\cite{Ta05}.    We note that there is also a very
small splitting expected between two bound levels,     $\frac{5}{2}^+$ and
$\frac{3}{2}^+$, at the excitation energy of $\sim$4.88 MeV. Our exploratory
calculation suggests a splitting of $\sim$160 keV  for both doublets.
Experimentally, the splitting of the $\frac{1}{2}^-$ and
$\frac{3}{2}^-$ states was found to be 152 $\pm$ 54 $\pm$ 36 keV.      The
splitting of the more bound positive-parity states has not been determined
quantitatively to date.

One can  extract the $\Lambda$-nucleus
spin-orbit   strength   from   the   splitting  of the $\frac{1}{2}^-$ and
$\frac{3}{2}^-$ states. Using  the  experimental  information  of  a 152
keV splitting for the two
odd-parity levels, we settle upon a $\Lambda$-nucleus spin-orbit strength
of 0.35 MeV.      We also conclude that the current knowledge of experimental
spectra is insufficient  to  assess any importance of the
$\Lambda$-nucleus $V_{sI}$ coupling in $^{13}_\Lambda $C.
For this reason, for the
$\frac{5}{2}^+$, $\frac{3}{2}^+$  level  splitting we  considered the
${}^9_\Lambda $Be   system  for  which  accurate experimental 
information  on   the
splitting of such levels has been obtained~\cite{Ha06}.
That information leads to a significant
constraint on the $\Lambda$-nucleus $V_{sI}$ coupling.

\section{Test of radiative-capture calculation with MCAS}
\label{Gamma-results}

\begin{figure}[h]
  \begin{center}
    \scalebox{.6}{\includegraphics{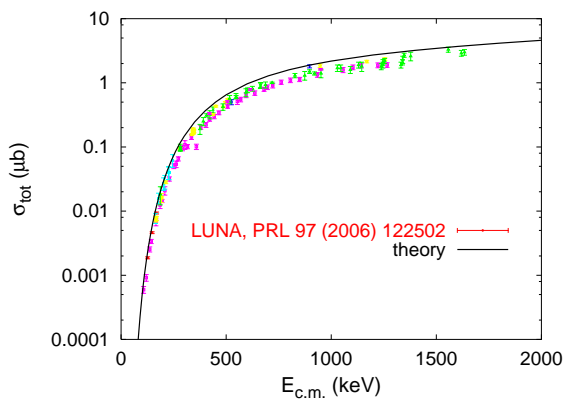}}
\caption{Total cross section of 
${^{3}{\rm He}}({\alpha},\gamma ){^{7}{\rm Be}}$
reaction.}
    \label{fig3}
  \end{center}
\end{figure}

In evaluation of radiative-capture reactions, we have to determine  the
(matrix-like) scattering wave function for bound and continuum states. 
In the case of the initial continuum state, this  can be rewritten in
closed form in terms of regular ($F_L$), irregular ($G_L)$,  and
outgoing ($O^{(+)}_L$) Coulomb functions, and a matrix-inversion
operation. This matrix inversion in the wavefunction expression retains all the
resonant and  nonresonant nuclear structures relevant for low-energy
capture and remove the need of formalisms (like R-matrix) which merely
parametrize known resonances.  The electro-magnetic (EM) operator of single-photon emission
is constructed via a generalization of the Siegert theorem, which
provides  the amplitude of the electromagnetic process in an explicitly
gauge-independent way.  A multipole decomposition of the relevant
Electric and Magnetic  transition matrix elements and the corresponding
results are given  in Refs.~\cite{CL,CLS}. Theoretical results are compared in
Fig.~\ref{fig3} with recent LUNA data as well as with other data sets (full
list is given in Ref.~\cite{CL}). Our results reproduce satisfactorily
the energy dependence but normalization appears overestimated ($\simeq$
1.4), and this represents a typical aspect of two-cluster model
calculations.

\section{Summary and Conclusions}
\label{conclusions}

We have applied the MCAS approach to study
the excitation spectra of light hypernuclei
$^9_\Lambda $Be and $^{13}_\Lambda $C.
The theoretical approach emphasises the single-particle motion of the
hyperon in the mean-field of the ordinary nuclear core, which is however
coupled to its own low-lying collective motions.

The phenomenological hyperon-core potential was constructed
from the ordinary nucleon-core potential by removing Pauli-blocking
effects, making a 15\% shrinkage of the radius, using
a $\frac{2}{3}$ reduction of the strength of the central potential,
and reducing drastically the spin-orbit potential.
However, the coupling of the single hyperon
with the collective motion of the core is
rather important in defining the hypernuclear spectra.
The deformation parameter is essentially that required
in the associated nucleon-nucleus dynamics.

From the recently observed fine splitting of the excited
hypernuclear spectra for these two nuclei, we conclude
that the  $\frac{1}{2}^-$-$\frac{3}{2}^-$ splitting in $^{13}_\Lambda $C
is dominated by the $\Lambda$-core spin-orbit interaction.
On the other hand, the fine splittings of the
$\frac{3}{2}^+$-$\frac{5}{2}^+$ doublet in both
$^9_\Lambda $Be and $^{13}_\Lambda $C are generated by
the coupling of the $s_\frac{1}{2}$-$\Lambda$ single-particle motion
with the $2^+$ collective excitation of the core.
As it involves an $s$-wave $\Lambda$-particle,
it is independent of the spin-orbit interaction. 
The recently measured structure of the fine
($\frac{3}{2}^+$-$\frac{5}{2}^+$) splitting in $^9_\Lambda $Be
enabled us to determine the sign and strength of that spin-spin
interaction, $V_{sI}$. 

We have also outlined the theoretical scheme to apply the MCAS  approach
to radiative capture, following Refs.~\cite{CL,CLS}.  We have considered in
particular the radiative-capture reaction  $\alpha$ + $^3$He $\rightarrow$
$^7$Be + $\gamma$, which can be reasonably described with a two-cluster
model. Our model calculation tests the feasibility of radiative-capture 
calculation within the Hamiltonian-type MCAS scheme and the
construction of EM operators that are explicitly gauge independent.

\end{document}